\documentclass[prb,preprint]{revtex4-1} 

\usepackage{amsmath,amsfonts,graphicx,amsbsy}
\usepackage[amssymb]{SIunits}
\usepackage{lineno}

\newcommand{\rmd}{\,\mathrm{d}}

\begin{document}
\title{``Meta'' relativity: Against special relativity?}
\author{J.~Rembieli\'nski}
\,\affiliation{Department of Theoretical Physics, University of Lodz\\
Pomorska 149/153, 90-236 {\L}{\'o}d{\'z}, Poland}
\email{jaremb@uni.lodz.pl} 
\author{M.~W{\l}odarczyk}
\affiliation{Department of Theoretical Physics, University of Lodz\\
Pomorska 149/153, 90-236 {\L}{\'o}d{\'z}, Poland}
\email{marta.wlodarczyk@gmail.com} 
\date{\today}

\begin{abstract}
We introduce a Lorentz-covariant description of tachyons,  free of inconsistencies. Our approach is based on an appropriate extension of the special relativity beyond the light barrier, owing to the freedom of synchronization of distant clocks.
\end{abstract}

\maketitle

\noindent
\emph{There was a young lady named Bright\\
Whose speed was far faster than  light...\\}

\section{Introduction}
In 1962, exactly fifty years ago, George Sudarshan and his collaborators,  Olexa Bilaniuk and Vijay Deshpande, published an article entitled ````Meta'' Relativity''\cite{meta}  in the American Journal of Physics. The authors introduced the notion of faster-than-light particles, which were subsequently named tachyons by Gerald Feinberg.\cite{feinberg} As was demonstrated in the Sudarshan's paper, tachyons have a number of counterintuitive properties. Sudarshan \textsl{et. al.} have shown that in particular
\begin{itemize} 	
    \item[--] velocity of tachyons is always greater than the velocity of light and can be arbitrarily large,
 	\item[--] energy of a tachyon increases with decreasing velocity, which means that tachyons accelerate as they lose energy,
 	\item[--] energy of a free tachyon can be negative.
\end{itemize}
Nevertheless, tachyon properties are in full agreement with the fundamental conservation laws of energy, momentum and  angular momentum.
Sudarshan's paper initiated a long-term dispute about the properties and possible role of tachyons in the classical and quantum physics. As a consequence very serious difficulties have been revealed in describing faster-than-light particles in the framework of the special theory of relativity. The most fundamental problem is inconsistency with the Einstein's meaning of causality: Causally related events involving tachyons are separated by space-like intervals, so time ordering of these events may not be invariant under the Lorentz transformations. This implies unavoidable violation of causality even if one applies the "reinterpretation principle", proposed  by Feinberg,\cite{feinberg} relying on reinterpretation of negative energy tachyons as antiparticles moving backward in time. This problem becomes even more serious at the quantum level as it leads to the breaking of the probability conservation in the quantum theory.\cite{Kamoi5,kam} Moreover, the velocity of a tachyon is not a well-defined quantity in the special relativity framework. This leads to the paradoxical notion of the so called ``transcendent tachyon''.\cite{sud_w_rec_5}  Furthermore, one cannot properly formulate the Cauchy initial conditions for tachyons.\cite{r2} Finally, the momentum  space of a free tachyon forms the  one-sheet hyperboloid,  which means that the tachyon energy  is not limited from below. At the quantum field theory level it leads, to the phenomenon of the quantum vacuum instability manifesting itself in a spontaneous creation of tachyon-antitachyon pairs from the vacuum.\cite{Kamoi5,kam}

Summarizing, it seems that  it is impossible to describe the tachyons consistently within the special relativity framework.	 Is there a way out of this seemingly dead-end situation?  The answer is yes, moreover the solution  lies in the special relativity itself. The key issue is the freedom of synchronization of distant clocks.

The paper is organized as follows: In Sec.~II we discuss the freedom of synchronization procedure in special relativity. In Sec.~III we introduce the absolute synchronization and the corresponding Lorentz group transformations. In Sec.~IV we formulate a Lorentz-covariant description of tachyons. Conclusions are given in Sec.~V.

\section{The freedom of synchronization}
As is well known, it is not possible to measure one-way (open path)  velocity of light without assuming a synchronization procedure for distant clocks. The issue and the meaning  of clock synchronization was elaborated in papers by Reichenbach,\cite{Rei} Gr\"unbaum,\cite{grunbaum} Winnie\cite{winnie1,winnie2} as well as in the test theories of special relativity by Robertson\cite{robertson} and  Mansouri and Sexl\cite{sexl1}
(see also Will\cite{will}).  A comprehensive discussion of the synchronization question is given by Lammerzahl\cite{lam} and Zhang.\cite{zhang1,zhang2} The point is that {only the average value of the speed of  light over closed paths is synchronization independent} (harmonic average over closed paths can be measurable with using one clock only). In other words, it is impossible to measure the one-way velocity of light without adopting a specific synchronization convention. Consequently, the  measured value of the one-way velocity of light is 
synchronization-dependent. 
In particular, the Einstein synchronization procedure, assuming the direction-independent speed of light, is only one (the simplest) possibility out of the variety of  possibilities which are all equivalent from the physical (operational) point of view.

\begin{figure}
    \centering
    \includegraphics[width=5in]{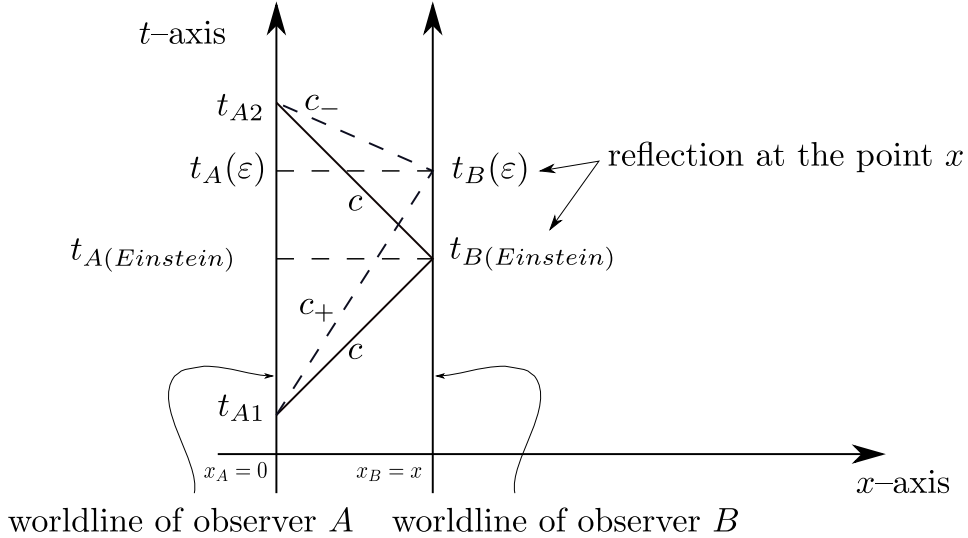}
    \caption{The Figure shows the synchronization procedure in Einstein (solid line) and general (dashed line) synchronization. In both cases the average speed of light over the closed path is the same and equals c.}
\label{synch}
\end{figure}

Intrasystemic synchronization of two clocks using light rays is schematically illustrated in Fig.~\ref{synch}.  At the time $t_{A1}$ observer $A$ located in at the point $x_A=0$ on the $x$-axis (worldline $A$) sends a light signal to observer $B$ located at the point $x_B=x$ (worldline $B$). The signal is reflected back to the observer $A$ at the point $x$ and arrives at the point $0$ at the time $t_{A2}$.
The speed of light from $A$ to $B$ is denoted by $c_+$ while from $B$ to $A$ by $c_-$. Because the path $ABA$ is closed, then the average light speed must be equal to $c$. From the kinematics (see Fig.~\ref{synch}) it follows that
\begin{equation}
\label{sredniac}
\left\langle|\rm{light\,velocity}|\right\rangle=\frac{2}{\frac{1}{c_+}+\frac{1}{c_-}}=c,
\end{equation}
i.~e. it is the harmonic average of $c_+$ and $c_-$. As the solution of Eq.~(\ref{sredniac}) we obtain
\begin{equation}
\label{c+-}
c_{\pm}=\frac{c}{1\pm \varepsilon},
\end{equation}
where the parameter $\varepsilon$ (synchronization coefficient) satisfies $-1< \varepsilon<1$ because of  $t_{A2}\geq t_B(\varepsilon)$ (see Fig.~\ref{synch}). For $\varepsilon=0$, $c_+=c_-=c$ so this case corresponds to the Einstein's choice (Einstein convention of synchronization).
One synchronizes clocks  assigning the reflection at $B$ the same time on the clocks of $A$ and $B$, namely
\begin{equation}\label{ta}
t_A (\varepsilon) = t_B(\varepsilon) := \frac{(1+ \varepsilon) t_{A1} + (1- \varepsilon) t_{A2} }{2},
\end{equation}
The synchronization coefficient $\varepsilon$ is more convenient for our purpose than the so called Reichenbach\cite{jammer}  coefficient $\varepsilon_R$, with which it is directly related by the formula:
$\varepsilon=1-2\varepsilon_R.$
Now, since
$x = c (t_{A2} - t_{A1})/2,$
we can relate the time
$
t_A (\varepsilon=0)\equiv t_{A (Einstein)}
$
  and
$
t_A(\varepsilon)$ by a formula $t_{A(Einstein)} = t_A (\varepsilon) +  \varepsilon x /c.
$

Extending the above considerations to three dimensions, one can relate the Einstein synchronization to an arbitrary one as follows:
\begin{equation}\label{zmiana_synchronizacji}
    t_E=t+\frac{\boldsymbol{\varepsilon}\boldsymbol{x}}{c}.
\end{equation}
Hereafter the subscript $E$ denotes quantities in the Einstein (i.e. standard) synchronization and  $t$ denotes the coordinate time in  a synchronization procedure defined by a vector coefficient $\boldsymbol{\varepsilon}$. Because the average speed of light over closed paths is the constant $c$, and by means of the causality requirement,  the  coefficient $\boldsymbol{\varepsilon}$  must be a vector inside a unit sphere  i.e. $\boldsymbol{\varepsilon}^2 < 1$.  In the following we restrict ourselves to a coordinate-independent synchronization coefficient  $\boldsymbol{\varepsilon}$. In that case the time redefinition (\ref{zmiana_synchronizacji}) is an affine transformation and therefore leaves the notion of inertial frame unchanged as well as it does not change the time lapse in a given point ($\rmd x = 0$ implies $\rmd t_E=\rmd t$ ). Of course, the position coordinate $\boldsymbol{x}$ and the underlaying Euclidean space geometry are synchronization-independent while,  for example, velocity is a synchronization-dependent notion. The same holds for the space-time metrics. In terms of the new coordinates the space - time geometry is related to the line element
\begin{equation}\label{invariant}
    \rmd s^2=g(\boldsymbol{\varepsilon})_{\mu\nu} \rmd x^{\mu}\rmd x^{\nu}
\end{equation}
with the coordinate-independent metric tensor
\begin{equation}
    \label{gepsilon}
    g(\boldsymbol{\varepsilon})=\left(
        \begin{array}{c|c}
        1&\boldsymbol{\varepsilon}^{\rm{T}}\\\hline
        \boldsymbol{\varepsilon}&-\delta_{ij}+\varepsilon^i \varepsilon^j
        \end{array}
    \right)
\end{equation}
where the superscript $\rm{T}$ denotes transpose of the column vector $\boldsymbol{\varepsilon}$. One can see that for the Einstein synchronization scheme ($\boldsymbol{\varepsilon}=\boldsymbol{0}$) the metric tensor reduces to the Minkowski form $g(\boldsymbol{0})=\rm{diag}(1,-1,-1,-1)$.
On the other hand the space geometry is given by the line element:\cite{Landau}
\begin{equation}
    \rmd l^2=\left(\frac{g_{0i}g_{0j}}{g_{00}}-g_{ij}\right)\rmd x^i\rmd x^j=\delta_{ij} \rmd x^i\rmd x^j=\rmd \boldsymbol{x}^2,
\end{equation}
so it is Euclidean in all inertial frames.

\begin{figure}
    \centering
    \includegraphics[width=3in]{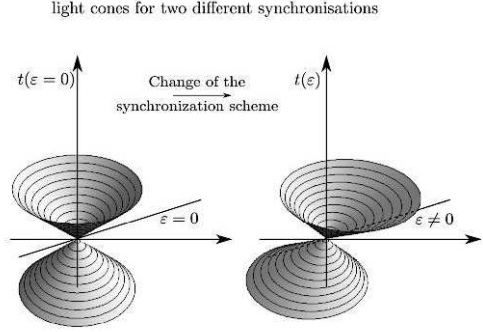}
    \caption{The figure shows the light cones for two different synchronizations (the first one being the Einstein synchronization). For  $\boldsymbol{\varepsilon}\neq 0$ the cone deforms.  }
\label{stozki}
\end{figure}

 From Eq.~(\ref{zmiana_synchronizacji}) it follows a relationship between velocities $\boldsymbol{v} =\rmd \boldsymbol{x}/\rmd t$  and  $\boldsymbol{v}_E =\rmd \boldsymbol{x}/\rmd t_E$  in the both synchronizations
\begin{subequations}\label{pr}
\begin{equation}
\label{predkosci}
    \boldsymbol{v}=\frac{\boldsymbol{v}_E}{1-\boldsymbol{\varepsilon}\boldsymbol{v}_E/c}.
\end{equation}
Eq.~(\ref{predkosci}) implies that one-way velocity of light in a direction $\boldsymbol{n}$ is given by
\begin{equation}\label{cn}
                                             \boldsymbol{c}(\boldsymbol{n}) =\frac{c \boldsymbol{n}}{1- \boldsymbol{\varepsilon}\boldsymbol{n}}                                                                    \end{equation}
\end{subequations}
i.e. it is 
anisotropic and synchronization-dependent (only in the Einstein synchronization it is direction-independent and always equal $c$).  However, this anisotropy is conventional as related to a specific clocks synchronization convention, unless the theory admits faster than light signals. As a consequence of Eq.(\ref{cn}), the light cone deforms under change of synchronization, which is illustrated in Fig.~\ref{stozki}.

Notice that Eq.~(\ref{predkosci}) is singular for superluminal velocities $\boldsymbol{v}_E$. This means that irrespectively of the apparent equivalence of the space-time geometry defined by the line element (\ref{invariant}) and the Minkowski one, the corresponding kinematics are nonequivalent for $|\boldsymbol{v}_E|>c$. The same statement is true for the causality notion. Indeed from (\ref{zmiana_synchronizacji}) it follows that
\begin{equation}\label{dtdt}
    \frac{\!\!\!\rmd t}{\rmd t_E}=1-\frac{\boldsymbol{\varepsilon}\boldsymbol{v}_E}{c},
\end{equation}
so the arrow of time can change for superluminal $\boldsymbol{v}_E$.

A crucial point is, how to use the synchronization freedom to solve the problem of describing tachyons. From the discussion above, it follows that the fundamental difficulty  lies in that the superluminal phenomena break the Einstein causality, which is the simple consequence of the Lorentz transformations changing the sign of the differential $\rmd t_E$. Therefore a question arises: {Is it possible to realize Lorentz symmetry in a way preserving sign of the coordinate time differential}?

\section{Lorentz group transformations in the absolute synchronization}
To answer the above question we follow the works of one of us.\cite{r1,r2,r3} Here we present a solution of this problem in a simple 1+1 dimensional case. The derivation is elementary.  We begin with the standard Lorentz transformations
\begin{subequations}\label{lorentz}
\begin{eqnarray}
\label{transformacjelorentza}
    t'_E&=&\frac{t_E- V_E x/c^2}{\sqrt{1-V_E^2/c^2}}, \\
    x'&=&\frac{x- V_E t_E}{\sqrt{1-V_E^2/c^2}}
\end{eqnarray}
\end{subequations}
where  $V_E$ denotes the velocity of the primed inertial reference frame $\Sigma'$, in the Einstein synchronization,  as seen by an observer in the unprimed frame $\Sigma$. Now, by means of Eqs. (\ref{zmiana_synchronizacji}), adapted to the 1+1 dimensional space-time, we eliminate in (\ref{lorentz}) the velocity $V_E$ and time $t_E$ in the Einstein synchronization and obtain

\begin{subequations}
\begin{eqnarray}\label{nowetransformacje}
    t'&=&\gamma(\varepsilon)\left[t\left(1+(\varepsilon+\varepsilon')\frac{V}{c}\right)+\frac{x}{c}\left(
    \varepsilon-\varepsilon'+(\varepsilon^2-1)\frac{V}{c}\right)\right],\\
\label{gamma}
    x'&=&\gamma(\varepsilon)\left(x-Vt\right),
\end{eqnarray}
with
\begin{equation}
    \gamma(\varepsilon)=\frac{1}{\sqrt{\left(1+\varepsilon \frac{V}{c}\right)^2-\left(\frac{V}{c}\right)^2}},
\end{equation}
\end{subequations}
where the velocity $V$ of the frame $\Sigma'$ and the Lorentz factor $\gamma(\varepsilon)$ are given  in the synchronization scheme $\varepsilon$ applied in the frame $\Sigma$ and because of (\ref{pr}) $-c_-<V<c_+$.  The synchronization coefficient applied in the frame $\Sigma'$ is denoted as $\varepsilon'$ . Now, to get transformations preserving the absolute causality requirement, the second term in the square bracket in Eq.~(\ref{nowetransformacje}) must vanish.  Fortunately, this last requirement simultaneously  implies the transformation law for the synchronization coefficient $\varepsilon$. As a result we obtain
\begin{subequations}\label{prim}
    \begin{eqnarray}
    \label{t}
        t'&=&\gamma(\varepsilon)^{-1}t\\
    \label{x}
        x'&=&\gamma(\varepsilon)\left(x-Vt\right)\\
    \label{epsilon}
        \varepsilon'&=&\varepsilon-\left(1-\varepsilon^2\right)\frac{V}{c}
    \end{eqnarray}
\end{subequations}
The above transformations form together a nonlinear realization of the Lorentz group, where nonlinearity takes place for the synchronization coefficient $\varepsilon$ only. This allows one to preserve the inertiality of frames. Now, the transformation rule for velocity $v =\rmd x/\rmd t$ obtained by means of Eq.~(\ref{prim}) is of the form
\begin{equation}\label{nowepredkosci}
    v'=\gamma(\varepsilon)^2(v-V)
\end{equation}
and is not singular for superluminal $v$, in contrary to the standard formula derived from Eqs.~(\ref{lorentz}).
The question arises: {What is the meaning of the synchronization coefficient $\varepsilon$}, which, as follows from (\ref{epsilon}), changes from frame to frame? Firstly, 
note that Eq.~(\ref{epsilon}) implies existence of an inertial frame where the synchronization coefficient is equal to zero. Thus in this distinguished (preferred) inertial frame $\Sigma_{PF}$ the standard Einstein synchronization applies. Now,  putting in Eq.~(\ref{epsilon})  $\varepsilon'= 0$ and denoting the velocity of  the preferred frame (as seen by the observer staying in the frame $\Sigma$) by $\vartheta$ we are able to express $\varepsilon$  as the function of $\vartheta$:
\begin{equation}
\label{epsilonodtheta}
    \varepsilon(\vartheta)=\frac{c}{2\vartheta}\left[\sqrt{1+\left(\frac{2\vartheta}{c}\right)^2}-1\right].
\end{equation}
Notice that $\varepsilon(0)=0$.  We also observe, that the velocity $\vartheta$ of the preferred frame $\Sigma_{PF}$, as seen by the observer  in the $\Sigma$, is related {via}~(\ref{nowepredkosci}) to the velocity $V_{\Sigma}$ of the frame $\Sigma$ as seen from the preferred frame $\Sigma_{PF}$ , by the formula    $\vartheta=-V_{\Sigma}/[1 - (V_{\Sigma}/c)^2 ]$. Therefore, from Eq.~(\ref{epsilonodtheta}) we have also a remarkable relationship between the synchronization coefficient $\varepsilon$ in the frame $\Sigma$ and the velocity $V_{\Sigma}:\quad \varepsilon = -V_{\Sigma}/c$. Note that the reciprocity principle does not hold in this synchronization scheme, i.~e. $\vartheta\neq -V_{\Sigma}$.

Now, taking into account Eq.~(\ref{epsilonodtheta}) we can eliminate the coefficient $\varepsilon(\vartheta)$ from the transformation rules (\ref{prim}) and~(\ref{nowepredkosci}). Furthermore, instead of~(\ref{epsilon}), we can use the  rule~(\ref{nowepredkosci}) adapted to the preferred frame velocity $\vartheta$. Consequently
\begin{subequations} \label{primm}
    \begin{eqnarray}
    \label{tt}
        t'&=&t \sqrt{[1+\varepsilon(\vartheta) V/c]^2-(V/c)^2},\notag\\&&\\
    \label{xx}
        x'&=&=\frac{x-Vt}{\sqrt{[1+\varepsilon(\vartheta) V/c]^2-(V/c)^2},}\notag\\&&\\
    \label{vartheta}
        \vartheta'&=&\frac{\vartheta-V}{[1+\varepsilon(\vartheta) V/c]^2-(V/c)^2}.\notag\\&&
    \end{eqnarray}
\end{subequations}
with  $\varepsilon(\vartheta)$ given by Eq.~(\ref{epsilonodtheta}).  Therefore, we have obtained nonlinear realization of the $1+1$ Lorentz group connecting the space-time coordinates of two arbitrary inertial frames and the velocities of the preferred frame as seen from those frames. Before rewriting the above transformations in a manifestly covariant way, some remarks are in order.

First of all, the inertial frames related by Eqs.(\ref{prim}) or (\ref{primm}) coincide with the standard lorentzian frames of reference. Indeed, by means of  Eqs.~(\ref{zmiana_synchronizacji} and \ref{prim})  applied to $t, t'$ and $V$ , we can return to the Einstein synchronization convention with standard (Lorentz) transformation laws. Thus the preferred frame, corresponding to $\vartheta=0$ (or equivalently $\varepsilon = 0$),  seems to be a formally distinguished frame only. Even the relativity principle can be formulated in the new language as: ``Each frame can be chosen as the preferred frame. Physics is unaffected by such a change''. However, this statements is true for luminal and subluminal phenomena only. Indeed, Eq.~(\ref{predkosci}), applied to velocities exceeding the   velocity of light, {are singular so irreversible}. Moreover, the notion of   Einstein causality 
in the case of superluminal objects does not coincide with the absolute causality related to the transformations (\ref{primm}), (see also (\ref{dtdt})). Therefore,  if we deal with superluminal phenomena, the preferred frame is not formally but physically distinguished, so the relativity principle is broken. Consequently,  the presented description of the space-time is equivalent to the special relativity if we take into account luminal and subluminal phenomena only. If superluminal phenomena take place, both descriptions, standard  and  the discussed above, are physically nonequivalent. The standard relativistic description is inadequate because of the difficulties mentioned in Section I. On the other hand, in the  absolute synchronization scheme described herein, the Lorentz symmetry and the causality notion survive also in case of the appearance of the superluminality, irrespectively of the breaking of the relativity principle.

\section{manifestly covariant formulation}
Now, the space-time geometry in the absolute synchronization scheme $\varepsilon(\vartheta)$ is described by the invariant line element of the form~(\ref{invariant})
\begin{equation}\label{invariantt}
    \rmd s^2=g(\vartheta)_{\mu\nu}\rmd x^{\mu}\rmd x^{\nu}
\end{equation}
where $x^0=ct$, $x^1=x$, $\mu,\nu=0,1$ and with the $\vartheta$-dependent metric tensor, changing from frame to frame according to Eqs.~(\ref{primm}), is given by
\begin{equation}\label{tensormetryczny}
    [g(\vartheta)_{\mu\nu}]=
    \left(\begin{array}{c|c}
        1&\varepsilon(\vartheta)\\\hline
        \varepsilon(\vartheta)&\varepsilon(\vartheta)^2-1
    \end{array}\right).
\end{equation}
Notice, that in the preferred frame $g(0)=\rm{diag}(1,-1)$, i.e.~we have the standard form of  the Minkowski geometry.

Now we are in a position to introduce a manifestly covariant description of the above formalism. To do this, let us define
\begin{subequations}\label{defczteropredkosci}
\begin{eqnarray}
    u^0&=&\sqrt{1-\varepsilon(\vartheta)^2},\\
    u^1&=&\frac{\varepsilon({\vartheta})}{\sqrt{1-\varepsilon(\vartheta)^2}},
\end{eqnarray}
\end{subequations}
and notice, that $u^1/u^0=\vartheta/c$ and $u^0 u^1=\varepsilon(\vartheta)$. In terms of $u^{\mu}$ the metric tensor can be written as
\begin{equation}
g(\vartheta)\equiv g(u)=\left(
                    \begin{array}{c|c}
                    1&u^0u^1\\\hline
                    u^0u^1&-(u^0)^2
                    \end{array}\right)\\
\end{equation}
and $g(u)_{\mu\nu} u^{\mu} u^{\nu}=1$.
The two-vector  $u^{\mu}$ forms the two-velocity of the preferred frame $\Sigma_{PF}$ as seen from the inertial frame $\Sigma$. Taking into account Eqs.~(\ref{epsilon},\ref{defczteropredkosci}) we get for  $u^{\mu}$ the transformation law exactly the same as for $x^{\mu}$ , given by (\ref{tt}) and (\ref{xx}).
 Thus the transformation rules (\ref{primm}) can be rewritten in a manifestly covariant form:
\begin{subequations}\label{kowariantne}
\begin{eqnarray}
    x'^{\mu}&=&D(V,u)^{\mu}_{\,\,\nu} x^{\nu},
    \\
    u'^{\mu}&=&D(V,u)^{\mu}_{\,\,\nu} u^{\nu}
\end{eqnarray}
\end{subequations}
where
\begin{subequations}\label{macierzrembielinskiego}
\begin{eqnarray}
    &&\!\!D(V,u)=\left(
        \begin{array}{c|c}
            \gamma(u)^{-1}&0\\\hline\vspace{0.3cm}
            -\frac{V}{c}\gamma(u)&\gamma(u)
        \end{array}
    \right), \\
    &&\!\!\gamma(u)\equiv\gamma(\varepsilon(u))=\frac{1}{\sqrt{\left(1+u^0u^1\tfrac{V}{c}\right)^2-\left(\tfrac{V}{c}\right)^2}}
\end{eqnarray}
\end{subequations}
The triangular form of $D(V,u)$   guarantees the absolute causality and simultaneity in this formalism. Moreover, the constant-time  hyperplane (here  $ \mathbb{R}^1$) is an invariant notion. This means that the time foliation of space-time is invariant under the Lorentz group action. Therefore, in contrast to the standard approach, we can formulate properly the Cauchy conditions in the case of  superluminal propagation.

\begin{figure}
    \centering
    \includegraphics[width=3in]{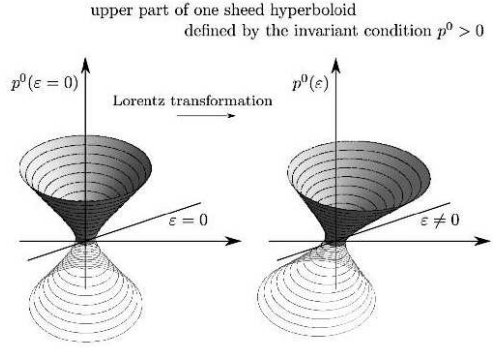}
    \caption{The figure shows the upper (physical) part of the one-sheet hyperboloid for tachyons for two values $\varepsilon=0$ and $\varepsilon\neq 0$ of the synchronization coefficient, related by the corresponding Lorentz transformation of $\varepsilon$ and momentum $p^{\mu}$. }
\label{tachion}
\end{figure}

\begin{figure}
    \centering
    \includegraphics[width=3in]{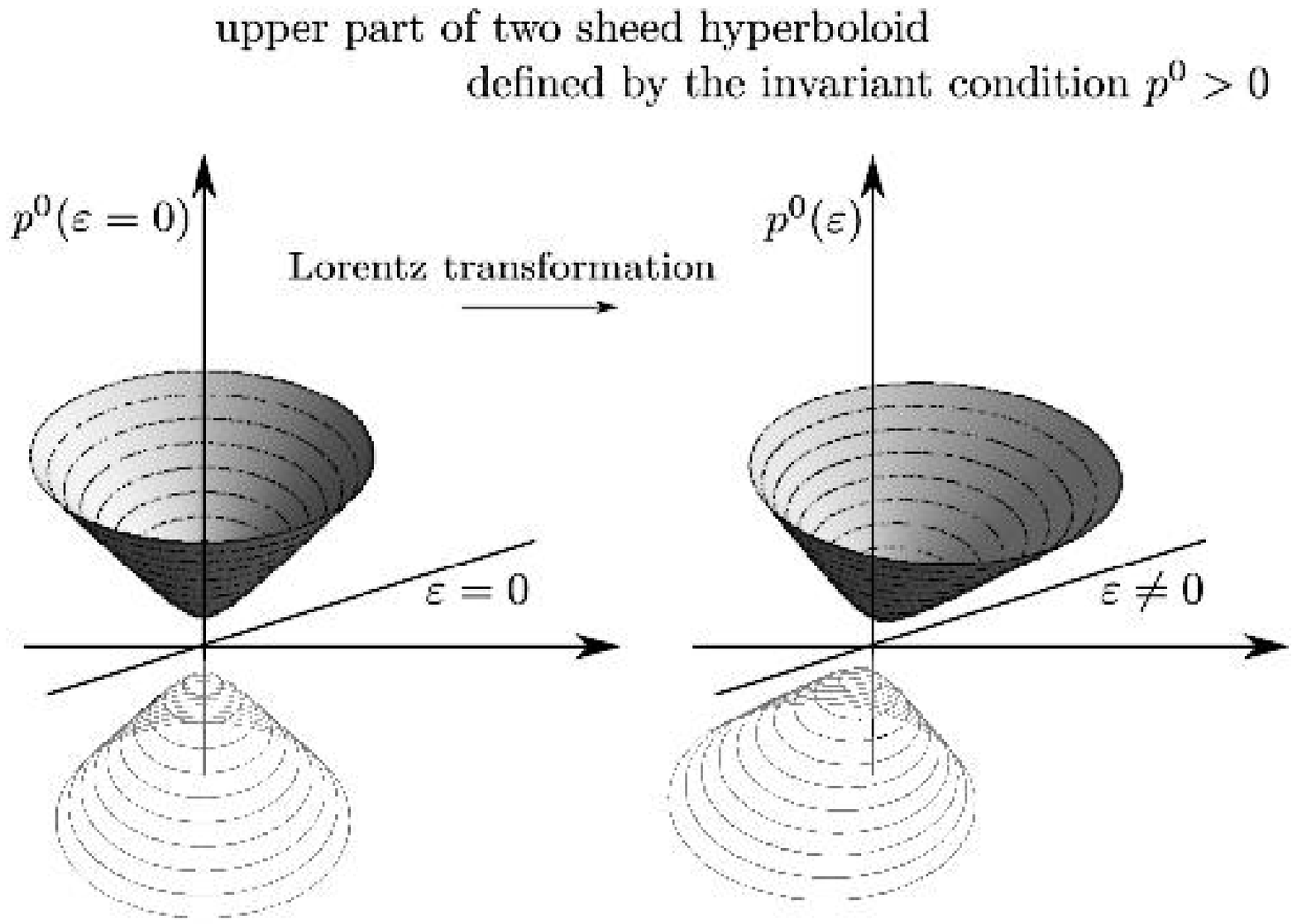}
    \caption{The figure shows the upper (physical) part of the two-shet hyperboloid for bradyons for two different values $\varepsilon=0$ and $\varepsilon\neq 0$ of the synchronization coefficient, related by the corresponding Lorentz transformation of $\varepsilon$ and momentum $p^{\mu}$.}
\label{bradion}
\end{figure}

As usual, we define the two-velocity $w^{\mu}:=\rmd x^{\mu}/\rmd \lambda$, where $\rmd \lambda=\sqrt{\left|g(u)_{\mu\nu}\rmd x^{\mu}\rmd x^{\nu}\right|}$,
satisfying   $g(u)_{\mu\nu} w^{\mu} w^{\nu}=\pm 1$ for subluminal $(+)$  and superluminal $(-)$ propagation,  respectively. By means of $w^{\mu}$ we can define momentum as $p^{\mu}= mcw^{\mu}$, obeying the dispersion relation
\begin{equation}\label{dyspersja}
    p^{\mu}p_{\mu}=\pm mc^2
\end{equation}
for ordinary particles (bradyons) and tachyons respectively. The  energy is identified with the (covariant) time translation generator i.e.~$E =c p_0$ . It is easy to notice that
$E = E_{Einstein}$  i.e. the energy is synchronization-independent.  A very important feature of the formalism is invariance of sign of the time component of the contravariant momentum irrespectively of the dispersion relation satisfied by $p^{\mu}$. Indeed, from Eqs. (\ref{kowariantne}) and (\ref{macierzrembielinskiego}) it follows that the sign of  $p^0$  is invariant under the modified Lorentz transformations. Thus the condition  $p^0 >0$ is invariant both for bradyons and tachyons. This means that the upper parts ($p^0>0$) of the  two sheet (bradyons) and one sheet (tachyons) hyperboloid $ p^{\mu}p_{\mu}=\pm mc^2$  are  carrying  spaces  of the Lorentz group both for bradyons and tachyons (see Figs.~{\ref{tachion} and \ref{bradion}}).
This fact is extremely important because it guarantees for tachyons the vacuum stability on the quantum level. Recall that in  the standard approach the tachyonic vacuum is unstable, which is frequently used as one of arguments against tachyons. The tachyon kinematics in the preferred frame remains exactly the same as for positive energy tachyons described in the fundamental paper by George Sudarshan.

\section{Conclusions}
	In the absolute synchronization framework all difficulties with description of tachyons disappear at last on the classical level. In particular there occur no causal paradoxes or transcendent tachyons, the tachyon velocity is well defined in all inertial frames of reference, the Cauchy conditions can be formulated for superluminal propagation and the Lorentz symmetry is preserved. However, the relativity principle is broken - the preferred frame is physically distinguished.   In the tachyonic sector there is no equivalence between the absolute and the Einstein formulation. On the other hand in the ordinary particles sector, both formulations are physically equivalent. This sector completely does not feel the existence of the preferred frame. Thus we can treat the approach presented here as a proper generalization of the special relativity beyond the velocity of light limit.
	A potential application of the introduced formalism  may be the attempt of reconciliation nonlocality of quantum mechanics with the Lorentz symmetry. It seems, that this generalization of the special relativity can be also helpful in a reformulation of the relativistic quantum mechanics that would solve some old standing  problems such as the problem of localization, non-uniqueness of the spin observable and others. The 3+1-dimensional formulation of the above approach was introduced  in the papers.\cite{r1,r2,r3} It was  also applied to the localization problem in Lorentz-covariant quantum mechanics\cite{cabremb}, description of the  Einstein- Podolsky- Rosen correlations\cite{rembsmol} and formulation of the Lorentz-covariant classical and quantum statistical physics\cite{kowrembsmol,nowe}. An idea of a correlation experiment identifying the preferred frame was given in the Ref.~\onlinecite{rembsmolepl}.
	
Finally, we conclude the paper by paraphrasing the second half of  the  Reginald Buller's  limerick, which first part was a motto of the article by George Sudarshan \textsl{et al.}\cite{meta} and this one. Namely instead of the original

\textsl{She set out one day,}

\textsl{In \textbf{a relative} way,}

\textsl{And returned home \textbf{the previous} night.
}

\noindent
we would rather prefer:

\textsl{She set out one day,}

\textsl{In \textbf{an absolute} way,}

\textsl{And returned home \textbf{the next} night.
}

\end{document}